\documentclass{ws-ijmpd}
\usepackage[compress]{cite}

\newcommand\neededonlyforijmpd[2]{#1}

\def\withonethinspace#1{\,\hbox{#1}}
\def\kms{\withonethinspace{km}\withonethinspace{s}^{-1}}
\def\kmsMpc{\kms\withonethinspace{Mpc}^{-1}}
\def\lsim{\mathop{\hbox{${\lower3.8pt\hbox{$<$}}\atop{\raise0.2pt\hbox{$\sim$}}
$}}}
\def\gsim{\mathop{\hbox{${\lower3.8pt\hbox{$>$}}\atop{\raise0.2pt\hbox{$
\sim$}}$}}} \def\goesas{\mathop{\sim}\limits}
\def\ave#1{\left\langle{#1}\right\rangle}

\usepackage{color}
\definecolor{MyGreen}{rgb}{0.21,0.64,0.18}
\definecolor{MyCyan}{rgb}{0.0,0.6,0.6}
\definecolor{MyDarkRed}{rgb}{0.71,0.14,0.07}

\newcommand{\edit}[1]{\color{MyDarkRed}{#1 }\color{black}}

\providecommand\hGpc{\mbox{$\,h^{-1}\,$Gpc}}
\providecommand\hMpc{\mbox{$\,h^{-1}\,$Mpc}}

\neededonlyforijmpd{
  
}{
  
}

\providecommand\muK{\mu\mathrm{K}}
\providecommand\dLum{d_{\mathrm L}}

\providecommand\jcap{Journ.~Cosm.~Astr.~Phys.\ }
\providecommand\apjl{Astrophys.~J.~Lett.\ }                 
\providecommand\apjs{Astrophys.~J.~Supp.\ }                 
\providecommand\apj{Astrophys.~J.\ }                 
\providecommand\aj{Astron.~J.\ }                 
\providecommand\prd{Phys.~Rev.~D}
\providecommand\PRL{Phys.~Rev.~Lett.\ }
\providecommand\mnras{Mon. Not. Roy. Astr. Soc.} 
\providecommand\aap{Astron.~Astroph.\ }

\providecommand\pasa{Proc.~Astr.~Soc.~Austr.\ }
\providecommand\cqg{Class.~Quant.~Gravit.\ }   %
\providecommand\grg{Gen.~Rel.~Gravit.\ }
\providecommand\nat{Nature}
\providecommand\jetpL{J.~Exper.~Theor.~Phys.~Lett.\ }

\providecommand{\refpositionsupvsnum}[2]{#2}

\newcommand\Ommzero{\Omega_{\mathrm{m0}}}
\newcommand\OmegakClarkson{\widehat{\Omega}_k}
\providecommand\MgII{Mg~{\sc II}}

\neededonlyforijmpd{
  \usepackage[dvips,breaklinks]{hyperref}
  \hypersetup{colorlinks,urlcolor=black,citecolor=black,linkcolor=black,filecolor=black}
  \usepackage{breakurl}
}{
  \providecommand\href[2]{\url{#1}} 
}

\begin{document}
\hypersetup{pdfpagelabels=false,breaklinks=false,linktocpage,citecolor=blue,linkcolor=magenta,filecolor=MyCyan,urlcolor=MyCyan}

\markboth{Buchert, Coley, Kleinert, Roukema \& Wiltshire}
         {Observational Challenges for FLRW}

\title{Observational challenges for the standard FLRW model}

\author{Thomas Buchert$^1$, Alan A. Coley$^2$, Hagen Kleinert$^3$, Boudewijn F. Roukema$^{4,1}$\\ and David L. Wiltshire$^5$}

\address{ $^1$Universit\'e de Lyon, Observatoire de Lyon,
Centre de Recherche Astrophysique de Lyon,\\ CNRS UMR 5574: Universit\'e Lyon~1 and \'Ecole Normale Sup\'erieure de Lyon,\\
9 avenue Charles Andr\'e, F--69230 Saint--Genis--Laval, France\footnote{BFR: During invited lectureship.}\\
$^{2}$Department of Mathematics and Statistics, Dalhousie University,\\ Halifax, NS B3H 3J5, Canada\\
$^{3}$Institute for Theoretical Physics, Freie Universit\"at Berlin, \\Arnimallee 14, 14195 Berlin, Germany\\
$^{4}$Toru\'n Centre for Astronomy, Faculty of Physics, Astronomy and Informatics, Grudziadzka 5,\\ Nicolaus Copernicus University, ul. Gagarina 11, 87--100 Toru\'n, Poland\\
$^{5}$Department of Physics and Astronomy, University of Canterbury,\\ Private Bag 4800, Christchurch 8140, New Zealand }

\neededonlyforijmpd{
  \maketitle

}{
}

\begin{abstract}
  {We summarise some of the main observational
  challenges for the standard
  Friedmann--Lema\^{\i}tre--Robertson--Walker cosmological
  model and describe how
  results recently
  presented in the
  parallel session ``Large--scale Structure and Statistics'' (DE3) at the
  ``Fourteenth Marcel Grossman Meeting on General Relativity''
  are related to these
  challenges.}
\end{abstract}


\neededonlyforijmpd{
  \ccode{PACS numbers: 98.80.-k, 98.80.Es, 98.80.Jk, 95.36.+x, 04.20.-q, 04.40.-b}
}{
}

\vspace{-4pt}

\section{Observational Challenges for the $\Lambda$CDM Model}\label{lcdmtrouble}

Despite the many well-known successes of the FLRW
model with its standard parameter values, henceforth denoted the $\Lambda$CDM model, there is a
wide range of observations with which it significantly
disagrees. The statistical significance of these
disagreements is very often debated from a Bayesian
perspective. If the $\Lambda$CDM is accepted as being
consistent with general relativity, then
one must contend with {\em a
  posteriori} statistics, also called the
look elsewhere effect, which globally
requires a \v{S}id\`ak-Bonferonni correction \cite{Abdi07}
for assessing overall statistical significance.
On the other hand, interpretation of structure formation
within the $\Lambda$CDM model is to a large degree based
on Newtonian physics---$N$-body simulations are widely seen as providing
state-of-the-art ways of comparing the FLRW model to observational
catalogues---but in comparison to general relativity,
the former should be assigned an extremely weak
prior.\footnote{For example, Keplerian orbits are disfavoured in relation to general-relativistic orbits at a
significance level of more
\refpositionsupvsnum{than\cite{WeisbNiceTaylor10} 100$\sigma$}
{than 100$\sigma$ \cite{WeisbNiceTaylor10}} based on the periastron decay of the
Hulse--Taylor \refpositionsupvsnum{pulsar \cite{HulseTaylor75} B1913$+$16}
{pulsar B1913$+$16 \cite{HulseTaylor75}}
(See Fig.~2 of Ref.~\refcite{WeisbNiceTaylor10}.)}
Although precise tests of general relativity have led to overwhelming Bayesian evidence vis-\`a-vis Newtonian gravity for binary systems on small scales, it could be argued that one cannot extrapolate any Bayesian comparison of the two theories to cosmological scales. However, there are no fundamental reasons to prefer Newtonian gravity over full general relativity on such scales, given its intrinsic theoretical shortcomings of absolute space and time and action-at-a-distance, redundant concepts which are more problematic as the scales grow larger. We will not attempt a full Bayesian analysis, which would require us to assign relative merits to full general relativity versus the FLRW model plus inhomogeneities obeying Newtonian gravity. Rather we will focus on individual observational constraints.

One observational contradiction with $\Lambda$CDM that is widely accepted in the
community is the primordial lithium abundance anomaly: the
lack of ${}^7$Li/H in metal-poor halo field stars in our
Galaxy is inconsistent with the $\Lambda$CDM cosmic
microwave background (CMB) expectation at
\refpositionsupvsnum{about\cite{Cyburt08Li7} $5.3\sigma$,}{about $5.3\sigma$ \cite{Cyburt08Li7},} 
unless new particles such as decaying
gravitinos are
\refpositionsupvsnum{assumed.\cite{Cyburt13Li7gravitino}}{assumed \cite{Cyburt13Li7gravitino}.}

While other difficulties for the $\Lambda$CDM model from the
CMB---the large-angle anomalies---are more often described
as ``tensions'', these disagreements with the model are numerous and their
statistical significance has tended to increase or remain stable as the accuracy and
precision of the data have \refpositionsupvsnum{increased.\cite{PlanckXXIIIisotropy,Planck2015isotropy}}{increased \cite{PlanckXXIIIisotropy,Planck2015isotropy}.}
At the largest angular scales
there is a lack of power, which translates to a {\em lack}
of observed structure in the CMB with respect to
$\Lambda$CDM on the largest scales. This was suspected in
the COsmic microwave Background Explorer (COBE) maps,
detected by the Wilkinson Microwave Anisotropy Probe (WMAP), and
confirmed by the Planck Surveyor first data
\refpositionsupvsnum{release.\cite{WMAPSpergel,Copi07,Gruppuso14Shalf,Copi13}}
{\cite{WMAPSpergel,Copi07,Gruppuso14Shalf,Copi13}.}
This observation means that the observed Universe is too
(spatially) {\em
  homogeneous} on the largest ($>10{\hGpc}$) scales, as
measured by $S_{1/2}$, the integral of the squared
auto-correlation of cosmological CMB
temperature--temperature fluctuations on angular scales
greater than 60$^\circ$ or spatial scales greater than the
radius to the surface of last scattering.\cite{Copi13}
One of the proposed explanations is
  that a finite universe model (without a spatial boundary)
  naturally explains
  this lack of \refpositionsupvsnum{power.
  \cite{Star93,Stevens93,
    WMAPSpergel,Aurich08a,
    LumNat03,RBSG08,RK11,RFKB13,
    AurichL2013WMAP,
    BielewPogosJaffePlanck13}}{power
  \cite{Star93,Stevens93,
    WMAPSpergel,Aurich08a,
    LumNat03,RBSG08,RK11,RFKB13,
    AurichL2013WMAP,
    BielewPogosJaffePlanck13}.}
In terms of spherical harmonic
decomposition of the sky maps, this is not restricted to
only the quadrupole
\refpositionsupvsnum{signal.\cite{Copi07,Gruppuso14Shalf,Copi13}} 
{signal \cite{Copi07,Gruppuso14Shalf,Copi13}.} 
Figure 3 and Table 2
in Ref.~\refcite{Copi13} show why the Planck data lead to
rejection of the $\Lambda$CDM model with probability
$p<0.0024$ for conservative versions of the data, or $p<
0.0003$ for the best quality data. Hemispherical asymmetry
rejecting the $\Lambda$CDM model at about the 3$\sigma$
level has remained a stable problem since it was first detected in
\refpositionsupvsnum{WMAP.
\cite{Eriksen04hemispheres,Hoftuft09,
  PlanckXVIcosmoparam13,
  MukherjeeADSS15}}
{WMAP
\cite{Eriksen04hemispheres,Hoftuft09,
  PlanckXVIcosmoparam13,
  MukherjeeADSS15}.}
Low spherical harmonic
mode number $l$ alignments in the CMB also remain
inconsistent with $\Lambda$CDM at about the 3$\sigma$ level
in the Planck
\refpositionsupvsnum{data. \cite{WMAPTegmarkFor,Copi13align}}
{data \cite{WMAPTegmarkFor,Copi13align}.}

Comparison of the integrated Sachs--Wolfe (ISW) effect in the CMB (WMAP5)
to stacked foreground voids, using luminous red galaxies (LRGs),
from the Sloan Digital Sky Survey Data Release~6 (SDSS DR6),
gives a temperature decrement
$ \Delta T_{\mathrm{ISW}} = 9.6\pm 2.2 \muK$, which is significantly
(at least 3$\sigma$) greater than the $\Lambda$CDM expected value of
\refpositionsupvsnum{$2.27 \pm 0.14 \muK$. \cite{Granett08ISW4sig,Flender13ISW3sig}}
{$2.27 \pm 0.14 \muK$ \cite{Granett08ISW4sig,Flender13ISW3sig}.}
Independently of the CMB,
Minkowski functionals of LRG-traced structure at scales of tens of megaparsecs in a volume of diameter 500{\hMpc}
in SDSS DR7 are also inconsistent with $\Lambda$CDM at
\refpositionsupvsnum{about\cite{WiegBO14} 3$\sigma$}{about 3$\sigma$ \cite{WiegBO14}}.

Combining many measurements at different scales, a 5$\sigma$ contradiction
is found between the standard $\Lambda$CDM model and observations, mainly
due to the differing power in the small-scale and large-scale parts of the
flat-space power
\refpositionsupvsnum{spectrum\cite{Battye14fivesigmaantiLCDM} $P(k)$}{spectrum $P(k)$ \cite{Battye14fivesigmaantiLCDM}}.
Comparison of low-redshift weak gravitational lensing of galaxies in the
Canada France Hawaii (Telescope) Lensing Survey (CFHTLenS) to CMB
constraints, interpreted according to $\Lambda$CDM, leads to
a present-day matter density parameter $\Omega_{\mathrm{m0}}$ which is too low
in the $\sigma_8$--$\Omega_{\mathrm{m0}}$ plane (where $\sigma_8$ is the
root-mean-square density fluctuation in a sphere of radius $8 h^{-1}$Mpc and
the Hubble constant is $H_0=100\,h\,\kmsMpc$), at a rejection level of $p\le
0.1$, unless sterile neutrinos are added to the
\refpositionsupvsnum{model. \cite{MacCrannJain14}}
{model \cite{MacCrannJain14}.}

The Baryon Acoustic Oscillation (BAO) signal of
Lyman~$\alpha$ forest absorbers in front of quasars
measured in the SDSS III/Baryon Oscillation Spectroscopic Survey (BOSS)/Data Release 11
contradicts the $\Lambda$CDM expectation at
\refpositionsupvsnum{about \cite{Delubac14BAOLyaforest} 2.5$\sigma$}{about 2.5$\sigma$ \cite{Delubac14BAOLyaforest}}.
The normalised growth rate dependence on redshift $f\sigma_8(z)$, as measured
in BOSS and the WiggleZ Dark Energy Survey (WiggleZ) \cite{Blake12WiggleZjoint}
is about $2\sigma$ too low in comparison to the Planck $\Lambda$CDM
expectation. (See Fig.~6 in Ref.~\refcite{ChuangBOSSLCDMruledout14}.)
A Bayesian multi-survey analysis comparing CMB, supernovae type Ia data and galaxy
surveys finds that the probability of
the $\Lambda$CDM model being correct
\refpositionsupvsnum{is \protect\cite{SalvatelliBruniWands14antiLCDM99} $p \le 0.01$}{is $p \le 0.01$ \protect\cite{SalvatelliBruniWands14antiLCDM99}}.

There may be a $\Lambda$CDM inconsistency with an overabundance of
luminous star-forming galaxies of $10^8 M_{\odot}$ forming too early,
at $z \approx 10$, as indicated by the number of Lyman break galaxies
(LBGs) detected so far using the Hubble and Spitzer Space
\refpositionsupvsnum{Telescopes. \cite{Melia14LBGstooearly}}
{Telescopes \cite{Melia14LBGstooearly}.}
Furthermore, the dark matter side of $N$-body
simulations in the $\Lambda$CDM cosmology is not able to readily account for
a number of the detailed features observed in the Local Group of
\refpositionsupvsnum{galaxies, \cite{Kroupa12DMcrisis}}
{galaxies \cite{Kroupa12DMcrisis},}
including the numbers, densities and spatial distribution of
dwarf satellite galaxies, in addition to a general overabundance of halos
in simulations as compared with what is observed.

\section{Observational tests of the FLRW geometry}

Many results of the standard cosmology are based on very large $N$-body numerical simulations using Newtonian gravity on a FLRW background. While such simulations yield values of the Newtonian potential in the present Universe no larger than $10^{-4}$, they presuppose the validity of the FLRW geometry. Model-independent observational tests of the validity of the FLRW geometry are therefore particularly important.

Such tests can be split into two classes: (i) tests of the validity of the Friedmann equations over long epochs of cosmic history in the regime where average Hubble expansion variation is nonlinear; (ii) tests of the validity of the spatial FLRW geometry below the scale of statistical homogeneity.
(Hereafter, as in common usage,
when stating ``homogeneity'', we refer to {\em spatial} homogeneity, unless otherwise
stated.)

\subsection{Tests of the Friedmann equation}
\nobreak
The violation of the FLRW relationship between the average expansion rate and
the luminosity distance can be used as a test of the importance of
\refpositionsupvsnum{inhomogeneities. \cite{ClarksonBL08, Larena09template, Wiltshire09timescape, Rasanen10propagstatdust, ShafielooClarkson10, Rasanen10propagstatgeneral, BoehmRasan13, LRasSzybka13, Wiltshire14, Chiesa14Larenatest}}
{inhomogeneities \cite{ClarksonBL08, Larena09template, Wiltshire09timescape, Rasanen10propagstatdust, ShafielooClarkson10, Rasanen10propagstatgeneral, BoehmRasan13, LRasSzybka13, Wiltshire14, Chiesa14Larenatest}.}
This is exemplified by the Clarkson--Bassett--Lu (CBL)
\refpositionsupvsnum{test: \cite{ClarksonBL08}}
test \cite{ClarksonBL08}:
the spatial
curvature parameter of the FLRW model, a constant, may be written as
\begin{equation}
  \OmegakClarkson ={[H(z)D'(z)]^2-1\over[H_0 D(z)]^2}\label{ctest1}
\end{equation}
for all redshifts, where $H(z)$ is the Hubble parameter, $H_0=H(0)$ the
Hubble constant, and $D(z)=\dLum/(1+z)$ where $\dLum$ is the luminosity distance
to a redshift $z$.
This test assumes that $H$ and $D$ depend on $z$ alone.
A further derivative of (\ref{ctest1}) with respect to
$z$ then gives a statistic which must be identically zero for all redshifts
if the Friedmann equations apply, irrespective of any dark energy model or
alternative model parameters. Model--independent observations of $H(z)$, $D(z)$
and $D'(z)$ for sufficiently large data sets can therefore be a powerful
discriminator of the standard cosmology from inhomogeneous cosmological models,
such as those with backreaction from inhomogeneities or exact inhomogeneous solutions.
Inhomogeneous cosmological models do not, in general, obey the
Friedmann equation at late epochs and do not have a uniform spatial curvature,
so that (\ref{ctest1}) is not constant with redshift. Predictions have been
made for specific models which incorporate
\refpositionsupvsnum{backreaction. \cite{Larena09template,Wiltshire09timescape,Wiltshire14,LRasSzybka13}}
{backreaction \cite{Larena09template,Wiltshire09timescape,Wiltshire14,LRasSzybka13}.}

\subsection{The scale of statistical homogeneity}

Many debates about the role of inhomogeneities in observational
challenges to the standard model involve different approaches to
the definition of a {\em scale of
statistical homogeneity}. Generally one must deal with spatial averages of
the density field, defined on a compact domain of a spatial hypersurface,
$\Sigma_t$, according to
\begin{equation}
\ave{\rho(t)}{_{{\cal D}_R}}=\frac1{{\cal V}(t)}\left(\int_{{\cal D}_R}{\rm d}
^3x\sqrt{\det{}^3\!g}\,\rho(t,{\mathbf x})\right),
\end{equation}
where ${\cal V}(t)\equiv \alpha R^3(t)=\int_{{\cal D}_R}{\rm d}^3x\sqrt{\det{}^3\!g}$ is the volume of
the domain ${\cal D}_R\subset\Sigma_t$, $g_{ij}$, ($1\le i,j \le3$) is the
intrinsic metric on $\Sigma_t$ and $\alpha$ is a dimensionless constant determined
by a choice of geometry; e.g., $\alpha=4\pi/3$ for Euclidean spheres. In approaches
in which ergodicity is assumed to apply, a definition of homogeneity often presupposes
the existence of an average positive density,
$\rho_0(t)$, defined by the limit
\begin{equation}
\lim_{R(t)\to\infty}\ave{\rho(t)}{_{{\cal D}_R}}=\rho_0(t)>0.
\end{equation}
A homogeneity scale, $\lambda_0(t)$, is then defined \cite{Gab05} by the
requirement that every point in $\Sigma_t$ be contained in a domain
${\cal D}_{\lambda_0}\subset{\cal D}_R$ such that
\begin{equation}
\left|\ave{\rho(t)}_{{\cal D}_R}-\rho_0(t)\right|<\rho_0(t)\qquad
\forall\; R>\lambda_0.
\label{homscale}\end{equation}
In practice, the density field can only be inferred indirectly from the
statistical properties of the distribution of galaxies, with all
of the systematic issues related to finite sample volumes and observational
biases. Thus, any practical measure of statistical homogeneity is not directly
based on a relation such as (\ref{homscale}), but rather on the
scale dependence of galaxy--galaxy correlation functions.

Observationally, the range 70--120$\hMpc$ comprises the smallest scales at which any notion of statistical
homogeneity can be argued to emerge \cite{HoggEis05,Scrimgeour12WDEShomog} at the
present epoch, based on
the 2-point correlation function. However, if all $N$-point correlations of
the galaxy distribution are taken into account, then the homogeneity
scale is expected to be reached, if at all, only on scales beyond
\refpositionsupvsnum{700$\hMpc$. \cite{WiegBO14}}
{700$\hMpc$ \cite{WiegBO14}.}
Improved survey quality (distance, volume and methods) tends to have increased
estimates of the homogeneity scale over the last few decades, e.g., see Ref.~\refcite{Kerscher01Mink}.
Another example is reported in Refs.~\refcite{RBOF15,RBFO15}: the $\approx105
h^{-1}$Mpc BAO scale, normally thought of as a comoving standard ruler, is
shrunk from about 6\% to 10\% across SDSS DR7 superclusters,
as a function of increasing overlap between luminous red galaxy (LRG) pair paths and
superclusters (assuming a $\Lambda$CDM
model when interpreting the observations).
Writing the metric near a supercluster using one of the standard FLRW expressions,
but replacing homogeneous parameters by effective parameters, this represents
a deviation in the spatial metric coefficients, $g_{ij}$,  of 6--10\%.

Estimates of the present epoch variation of the density on the largest
possible scales in SDSS DR7, limited only by survey
volumes\footnote{Specifically, Sylos Labini {\em et al} \cite{sl09} divided the full
sample of 53,066 luminous red galaxies in the redshift range $10^{-4}<z <0.3$
into $N$ equal nonoverlapping volumes. The standard deviation of order 8\% is
found for the range $4\le N\le15$.}, have given
a standard deviation of order \cite{sl09} 8\%, consistent {with}
an earlier measurement of 7\% in
a smaller
\refpositionsupvsnum{sample. \cite{HoggEis05}}
{sample \cite{HoggEis05}.}
In the $\Lambda$CDM cosmology
the understanding of this standard deviation is subject to
the observational interpretation of cosmic variance arising from the
evolution of initial density perturbations, given that the assumption of
ergodicity will not hold on every assumed large scale. Hence, the amplitude of
(\ref{homscale}) on large scales does not itself provide a model--independent
test of the FLRW geometry.

\subsection{Tests of spatial geometry on ``small'' scales}

Even if one accepts the most conservative estimate of
70--120$\hMpc$ as the scale at which some notion of statistical
homogeneity emerges, with an effective FLRW geometry at
larger scales, then the assumption that such a geometry applies
at smaller scales is no more than a working
hypothesis\footnote{It is sometimes claimed that the FLRW
  metric is applicable on all scales, ``except in the
  immediate vicinity of black holes and
  \refpositionsupvsnum{neutron stars''. \cite{GW14howwell}}
  {neutron stars'' \cite{GW14howwell}.}
  However, such statements merely
  reflect the {\em assumptions} of the standard
  cosmological model, rather than direct
  observational facts or general mathematical theorems. In
  view of the observational results reviewed
  here, we know that the assumptions made in
  Refs.~\protect\refcite{GW14howwell,GW11newframework} are
  not applicable to the real physical Universe. A critique
  of the mathematical results of
  Refs.~\refcite{GW14howwell,GW11newframework} is given in
  Ref.~\refcite{GWdebunk15}, and is discussed further in
  Ref.~\protect\refcite{Ostrowski15MG14GW}.} which must be
seriously questioned given that the largest {\em typical}
inhomogeneities are voids of diameter $\goesas30\hMpc$ and
density contrast $\delta\rho/\rho\goesas-0.95$, which form some
40\% of the volume\footnote{The overall statistics \cite{Pan2012voids}
include a small fraction of larger voids and a large population
\refpositionsupvsnum{of smaller minivoids, \cite{minivoids}}
{of smaller minivoids \cite{minivoids},}
making the Universe void dominated
at the present epoch with typical theoretical
estimates of the void filling-fraction
\protect{\refpositionsupvsnum{of up to
\protect\cite{Colberg08voidcomparison,Cautun15voidfrac80pc}
80\%.}{of up to 80\%
\protect\cite{Colberg08voidcomparison,Cautun15voidfrac80pc}.}}
By volume the Universe is mostly empty, with a very spiky
density distribution in the remaining tiny volumes.} of the Universe at low
\refpositionsupvsnum{redshifts. \cite{HoyleVOg02voids,HoyleVog04voids}}
{redshifts \cite{HoyleVOg02voids,HoyleVog04voids}.}
Density variations
of order 100\% are thus guaranteed when smoothing on scales of tens of megaparsecs,
and constitute a regime which is accepted as nonlinear in the standard
cosmological model. This regime is commonly treated by
Newtonian $N$-body simulations, leading to many phenomenologically
realistic results. However, given that the regime is nonlinear then the first
principles of general relativity, which demand a coupling of matter and geometry,
have no {\em a priori} preference for the FLRW geometry modified by Newtonian gravity on these scales.
This is why there have been several recent attempts to construct relativistic
\refpositionsupvsnum{simulations,
\cite{AdamekDDK15,
  MertensGS15BSSNmethod,GiblinMS15departFLRW,
  BentivegnaBruni15}}
{simulations
\cite{AdamekDDK15,
  MertensGS15BSSNmethod,GiblinMS15departFLRW,
  BentivegnaBruni15},}
some of which are summarised in
Ref.~\refcite{Bruni16MG14}.

Furthermore, the hypothesis of applicability of the FLRW spatial geometry on scales $\lsim100\hMpc$ is open to direct observational test.
If the hypothesis were true, then all motions of galaxy clusters---the
largest bound structures---should reduce to a uniform FLRW expansion plus
local Lorentz boosts, in contrast to general inhomogeneous solutions of
Einstein's equations which exhibit a differential expansion of space.
If all departures from homogeneity are described by local boosts on an
FLRW geometry, then the dipole anisotropy in the CMB
is purely kinematic, as is conventionally assumed.

In addition to the known motions of the Sun in our galaxy,
and our galaxy's motion within the Local Group of galaxies,
the kinematic interpretation of the CMB dipole requires that
the local group is boosted at $635\pm38\kms$ in a direction
\refpositionsupvsnum{$(\ell,b)=(276.4^\circ,29.3^\circ)\pm3.2^\circ$. \cite{Tully08preHpec}} 
{$(\ell,b)=(276.4^\circ,29.3^\circ)\pm3.2^\circ$ \cite{Tully08preHpec}.}
However, decades of work by
astronomers to explain the amplitude of this ``clustering
dipole'' has led to an ongoing debate about the convergence
of bulk flows on
\refpositionsupvsnum{scales $\lsim120\hMpc$. \cite{LavauxTMC,BilickiChJM11}}
{scales $\lsim120\hMpc$ \cite{LavauxTMC,BilickiChJM11}.}
Very recently, a resolution
of this debate has been claimed in
Ref.~\refcite{HessKitaura14} on the basis of constrained
$N$-body simulations, which incorporate various technical
improvements such as estimates for uncertainties associated
with missing attractors. However, given the complexities of
the statistical interpretation of numerical simulations---e.g.,
the definition of what the peculiar velocity is on a
sample that is considerably smaller than the homogeneity
scale---and the 20\% uncertainty in the magnitude of the final result in
Ref.~\refcite{HessKitaura14}, it is likely that the debate
will continue.

In contrast, a recent purely observational model-independent
analysis \cite{Wiltshire12Hflow} of the COMPOSITE sample of
4,534 group and cluster distances found that the spherically
averaged Hubble expansion is very significantly more uniform
in the rest frame of the Local Group (LG) as compared to the
standard rest frame of the CMB\footnote{A further search for
  a rest frame of minimum nonlinear Hubble expansion
  variation \cite{McKayWilt15} finds that the LG frame is
  consistent with such a frame but the statistical
  likelihood is not changed significantly under local boosts
  within the plane of our galaxy on account of a lack of
  constraining data in the region obscured by the Milky
  Way. Such degenerate boost directions do not include that
  from the LG to CMB rest frame, however.} (The Bayesian
evidence is very strong, with $\ln B>5$.) While this result
is completely unexpected if the clustering dipole is purely
kinematic, the residual monopole and dipole variation of the
Hubble expansion in the Local Group frame was found to be
consistent with a 0.5\% anisotropy in the distance--redshift
relation on $\lsim65\hMpc$
\refpositionsupvsnum{scales, \cite{Wiltshire12Hflow}}
{scales \cite{Wiltshire12Hflow},}
an effect which can be reproduced using nonlinear exact
solutions of Einstein's equations with inhomogeneities no
larger than the scales in
\refpositionsupvsnum{question. \cite{BolNazWilt16}}
{question \cite{BolNazWilt16}.}

Although the Planck Surveyor team claims to have measured Lorentz boosting
through the effects of frequency modulation and aberration on the angular power
spectrum, the boost direction is consistent with that of the CMB dipole only
when small angular scales are
\refpositionsupvsnum{considered. \cite{PlanckXXVIIDoppBoost}}
{considered \cite{PlanckXXVIIDoppBoost}.}
When large angle
multipoles are included the dipole moves across the sky to point in the
modulation dipole anomaly
\refpositionsupvsnum{direction. \cite{Hoftuft09}}
{direction \cite{Hoftuft09}.}
The scale-dependence of
this result in Ref.~\refcite{PlanckXXVIIDoppBoost} therefore suggests an intriguing possibility that
certain large-angle anomalies in the CMB power spectrum may arise from
treating a partly nonkinematic dipole purely kinematically; a direct
result of the relevant geometry being non-FLRW. In an independent study,
the hypothesis of a purely kinematic origin for the dipole in the cosmic
distribution of radio galaxies has been rejected at the 99.5\% confidence
\refpositionsupvsnum{level, \cite{RubartSchwarz13}}
{level \cite{RubartSchwarz13},} using the NRAO VLA Sky Survey (NVSS).

\section{Recent developments presented at the MG14 meeting} \label{s-MG14-results-general}

Recent work in statistical assessment
of large-scale structure was presented at the
``Fourteenth Marcel Grossman Meeting on General
Relativity'' (MG14) parallel session DE3, ``Large--scale Structure
and Statistics''.  This included higher-order
statistics, morphological properties, the reality and
significance of large structures in the Universe, standard
rulers like the baryon acoustic oscillation (BAO) peak
location, comparisons with mock catalogues, path finders for
next-generation galaxy catalogues, non-Gaussian statistics,
mass functions and abundance statistics of collapsed
objects, among other topics.  Contributions also included
general-relativistic aspects of large-scale structure
formation in these contexts, such as the measurement of
backreaction in large-scale structure data, indirect
measures of metrical properties and improved
redshift--distance measures in inhomogeneous cosmologies.
The session was divided into observational issues and
modelling aspects. 
In Sect.~\ref{s-MG14-DE3-obs-results}
we mainly concentrate on the former,
by looking at various observational cornerstones and
highlighting current challenges to the standard
Friedmann--Lema\^{\i}tre--Robertson--Walker (FLRW) model and those
likely to be faced in the near future. Marco Bruni's report
briefly summarises modelling work that was contributed to
\refpositionsupvsnum{the parallel session. \cite{Bruni16MG14}}
the parallel session \cite{Bruni16MG14}.

Theoretical issues,
including backreaction and inhomogeneous cosmology, were
discussed in another MG14 parallel session, DE2.
Possible observational challenges to the
standard FLRW model\footnote{See Ref.~\refcite{Bull15BeyondLCDM} for another
recent summary of challenges to the standard model.}
and theoretical discussion of the
effects of inhomogeneities were discussed in
the individual talks of both the DE2 and DE3 sessions and during
a special discussion session. (See, in particular, the
summary of Ostrowski's presentation \cite{Ostrowski15MG14GW}.)
A statistical question that does not yet seem to have
been considered
in observational cosmology was also raised for discussion
in the context of calculating variances (e.g., when quantifying backreaction terms):
heavy tails in the distributions of physical quantities,
such as those of the orbits of strongly interacting
\refpositionsupvsnum{particles, \cite{Kleinert12particleLevy}\footnote{\protect\url{http://www.physik.fu-berlin.de/~kleinert/talks/CastTaiw1.pdf}}}
{particles \cite{Kleinert12particleLevy},\footnote{\protect\url{http://www.physik.fu-berlin.de/~kleinert/talks/CastTaiw1.pdf}}}
could lead to the need
to study heavy-tailed distributions such as
the L\'evy distribution, which has a finite
mode and median but an infinite mean and variance,
\cite{Podobnik00Levyflight} although truncated
forms (which could occur if the Universe has a finite spatial volume) might be more realistic.

\section{Recent observational results} \label{s-MG14-DE3-obs-results}

The DE3 parallel session at MG14 included recent work that
touches on several of the above
issues\footnote{\refpositionsupvsnum{\protect\url{http://www.icra.it/mg/mg14/parallel_sessions.htm}}{\protect\href{http://mg14reg.icra.it/mg14/FMPro?-db=3_talk_mg14_.fp5&-lay=talk_reg&ps::web_code=0555773014&-format=session_mg14.htm&main_1::Attivo=yes&talk_accept=yes&-SortField=order2&-SortOrder=ascend&-Max=50&-Find}{{\tt
      http://www.icra.it/mg/mg14/}};} archived:
  \url{http://www.webcitation.org/6c50Ydtut}}.  The
observational disagreements with $\Lambda$CDM listed above
indicate a problem in the {\em accuracy} of the model,
{i.e.,} they are concerned with both systematic and random
error, rather than a problem in the {\em precision} of the
model, which is concerned with random error.  Results
indicating systematic errors that are normally unaccounted
for in $\Lambda$CDM observational analyses, i.e., problems of
{\em accuracy}, include Shanks' presentation of the
``Local Hole'' on a 100--200{\hMpc} scale,
Bolejko's discussion of a general-relativistic study of
Hubble flow anisotropy on somewhat smaller scales, and
Roukema's explanation of the recently found flexibility of
the BAO scale. The other talks mostly presented observational
results or projects that could potentially contribute
to $\Lambda$CDM's falsifiability in the coming decades,
while not presently rejecting it.
These include Mackenzie's presentation on
the CMB Cold Spot and the ISW; Majerotto's
discussion of the CBL test; Pisani's
explanation of how to use 10--200{\hMpc} voids as a
cosmological probe; and Kaminker's
quasi-periodical statistical analysis of SDSS data.

A fundamental difficulty in observational cosmology follows
from elementary geometry in a space that is approximately
flat (or not too hyperbolic).  Within a few 100{\hMpc} from
the observer, the spatial section contains very few
$(100{\hMpc})^3$ volumes,
leading to a high Poisson error in
estimating any large-scale physical parameter, while at
distances of 1{\hGpc} or more, the Poisson noise
(for a 100{\hMpc}-scale statistic)
is more
reasonable.  This problem is commonly termed ``cosmic
variance''.
Recall that a general-relativistic interpretation of
nearly empty spatial regions assigns a negative
averaged curvature to these regions, which---if a Newtonian
point of view is taken---implies ignorance of a physical
variance in the curvature.
(The cosmic variance problem also occurs for larger-scale
statistics.) At the 100{\hMpc} scale, there are two basic
strategies: ignore the volume within a few 100{\hMpc} of the
observer and marginalise over a range of relativistically
valid models of this volume, or impose a single theoretical
model on the data analysis. Both approaches introduce {\em a
  priori} assumptions into the data analysis. The former
would in principle be better, but the latter is what is used
in practice.

Shanks presented new evidence that strengthens results
that were suspected from faint galaxy number counts in the
1980s and were seriously quantified in the 2000s using the
2 Micron All-sky Survey
\refpositionsupvsnum{(2MASS), \cite{Frith03}}
{(2MASS) \cite{Frith03},}
and by
photometric followup of bright galaxies in the Two Degree
Field Galaxy Redshift Survey
\refpositionsupvsnum{(2dFGRS). \cite{Busswell04}}
{(2dFGRS) \cite{Busswell04}.}
These tentatively identified a 150--300{\hMpc} scale
underdensity with respect to more distant galaxies,
including a celestial North--South asymmetry, dubbing it the
``Local Hole''.  Whitbourn and Shanks confirmed this
\cite{WhitbournShanks14} by analysing 250,000 galaxy
redshifts from the Six Degree Field Galaxy Survey (6dFGS)
and the SDSS, using both their redshift distributions and
number counts in comparison to the deeper Galaxy and Mass
Assembly (GAMA) survey. They also made peculiar velocity
maps that rejected the possibility of the local 150{\hMpc}
diameter region being at rest in the CMB frame at 4$\sigma$.
In as yet unpublished results together with Whitbourn,
Shanks explains how clustering-independent luminosity
function analysis, an independent galaxy survey, and a
complete X-ray cluster survey corroborate their earlier
results.

Studies of the pattern of galaxies' peculiar
velocities---defined by subtraction of either a na\"{\i}ve
(strictly linear) Hubble
\refpositionsupvsnum{law, \cite{DavisScrimgeour14naiveH}}
{law \cite{DavisScrimgeour14naiveH},}
or a third-order Taylor expansion or analytically exact
expression for the redshift--distance relation for an FLRW
model---have led to decades of community debate about
whether cosmic flows are compatible or incompatible with
$\Lambda$CDM
{e.g., Refs.~\refcite{LavauxTMC,Tomita00cosmicflows,Springob15bulk2MTF} and
references therein).
Bolejko's new work with Wiltshire and Nazer \cite{BolNazWilt16} provides a fresh look at this model-dependent observational debate. They have performed numerical simulations by ray-tracing in exact Szekeres models with inhomogeneous structures which match as closely as possible the dominant inhomogeneities observed on scales $\lsim80\hMpc$, while asymptotically matching a Planck--satellite normalized $\Lambda$CDM model on larger scales. These simulations are quantitatively constrained by the requirement that they be consistent with the observed CMB anisotropies, while incorporating a nonkinematic CMB dipole component. The dipole and quadrupole variations of the local Hubble expansion are then quantitatively compared to those of the COMPOSITE sample data as studied earlier in Ref.\ \refcite{Wiltshire12Hflow}. It is found that the Szekeres model is able to more closely model observational features than either a FLRW model with kinematic boost from the LG to CMB frame, or Newtonian $N$-body simulations. In fact, the latter models are rejected at more than the 2$\sigma$ level.

Either an increased observational confidence in the existence of the Local Hole or a preference for the Szekeres model over the standard model on a smaller scales would separately or jointly imply that systematic errors
need to be taken into account for many, though not all,
observational analyses on 1--20{\hGpc} scales. Angles and
distances to objects beyond 100--200{\hMpc} depend on
correctly propagating light through the geometry on scales inside of 100--200{\hMpc}.
This point is not always appreciated, with the incorrect perception
that inferences of FLRW cosmological parameters
from CMB observations are independent of the assumed foreground
geometry, whether this is on a tens of megaparsecs scale or greater.
The possibility of a nonkinematic contribution to the CMB dipole is
actually directly relevant to the question of large-angle anomalies, and
a future challenge is to investigate this directly in the CMB map--making
procedures.

On a scale up to a little less than 1{\hGpc}, Roukema
presented recent SDSS DR7 work with Buchert,
Fujii and Ostrowski \cite{RBFO15} extending previous
work \cite{RBOF15} that showed that the $\approx105${\hMpc}
BAO peak location, normally thought of as a comoving
standard ruler, is shrunk by a fraction of 6\% across SDSS
DR7 superclusters. This is expected from
scalar averaging, which can be seen
as a general-relativistically more careful generalisation of the FLRW
\refpositionsupvsnum{model, \cite{Buch00scalav,Buch01scalav,Buchert08status}}
{model \cite{Buch00scalav,Buch01scalav,Buchert08status},}
but has
not been predicted in the $\Lambda$CDM model.
The expected redshift-dependent shift (rather than an environment-dependent shift)
has been
\refpositionsupvsnum{modelled, \cite{Desjacques10BAObias,SherwZald12peakscale}}
{modelled \cite{Desjacques10BAObias,SherwZald12peakscale},}
but
the expected shift is tiny, e.g.,
\refpositionsupvsnum{$<0.3\%$. \cite{SherwZald12peakscale}}
{$<0.3\%$ \cite{SherwZald12peakscale}.}
If the environment dependence were genuinely an environment-dependent
effect, then it should strengthen as the required overlap between galaxy pairs
and superclusters increased. This is indeed the case, as was shown in
the recent work presented at the meeting, with the shift increasing up
to 10\% of the peak
\refpositionsupvsnum{location. \cite{RBFO15}}
{location\cite{RBFO15}.}
This presently only qualifies
as a qualitative inconsistency with $\Lambda$CDM. Future work should confirm
whether or not the environment-dependent BAO peak location shift is
quantitatively consistent with $\Lambda$CDM.

The other observational presentations at the DE3 session concerned ongoing
projects that
do not presently reject the $\Lambda$CDM model, but could potentially
help to separate the $\Lambda$CDM model from relativistic inhomogeneous models
and explicitly non-GR models.
Mackenzie presented work he is doing with Shanks
on the ISW effect by cross-correlating the LRG distribution
using photometric redshifts from a $\goesas$ 4700 deg$^2$ photometric southern optical survey
with Planck Surveyor CMB maps. He plans to test the nature
of the $\sim400${\hMpc} supervoid \cite{Szapudi14coldspotsupervoid} in front of
the CMB Cold Spot and whether the two are physically
related; see also Ref.~\refcite{RomanoCC15HLQGcmb}.

Majerotto presented work
together with Sapone and Nesseris
\cite{SaponeMN14}
who applied the Clarkson, Bassett and Lu test \cite{ClarksonBL08,Clarkson12test}
using eight $H(z)$ cosmic chronometer estimates \cite{Moresco12eightH}
(age of oldest passively evolving red galaxies
at any given \refpositionsupvsnum{redshift\edit{\cite{JimLoeb02cosmchronom}} $z$}{redshift $z$\cite{JimLoeb02cosmchronom}})
and a recent compilation of supernovae type Ia redshift--magnitude estimates.
This test should distinguish some classes of relativistic or other non-FLRW
models from the FLRW model.
The uncertainties with the presently used data sets were found to be large,
with the FLRW model being found to be consistent with the data.
What is particularly interesting for the coming decade are predictions
for redshift evolution of $\OmegakClarkson$,
the average curvature parameter, from
telescopes such as Euclid. Majerotto showed that
statistically homogeneous and isotropic general-relativistic
cosmological models in which the expansion rate history is observationally
realistic, including the Timescape \cite{Wiltshire09timescape,DuleyWilt13,NazerW15CMB}
and Tardis \cite{LRasSzybka13} models, have $\OmegakClarkson(z)$ relations
that should be observationally distinguishable from $\Lambda$CDM to high
significance by
\refpositionsupvsnum{{\em Euclid}. {\cite{SaponeMN14}}}
{{\em Euclid} {\cite{SaponeMN14}.}}
Both the models, as in the case of
template metrics that match the
supernovae type Ia distance-modulus--redshift relation,
i.e., that of Ref.~\refcite{Larena09template} and
the virialisation
\refpositionsupvsnum{approximation, \cite{ROB13}}
{approximation \cite{ROB13},}
have an average curvature
parameter which evolves from a small value to a strongly negative average
effective curvature today. In Ref.~\refcite{Larena09template} it is also concluded that this
curvature evolution is detectable by {\em Euclid}.

It has been known for at least three
decades \cite{deLappGH86}
that cosmic voids exist on scales of tens of megaparsecs.
These have recently become the subject of systematic
\refpositionsupvsnum{study, \cite{HoyleVOg02voids,HoyleVog04voids,Pan2012voids}}
{study \cite{HoyleVOg02voids,HoyleVog04voids,Pan2012voids},}
especially since the release of the
SDSS. Voids on scales ranging from $\sim$10--200{\hMpc} are
now the subject of intense debate, linked to the fundamental
disadvantage in detecting underdensities as opposed to
\refpositionsupvsnum{overdensities. \cite{NadHot2013, SutterLWWWP14,
  PisaniLSW14stack, Pisani15voidcounts,
  PisaniSW15vpec}}
{overdensities \cite{NadHot2013, SutterLWWWP14,
  PisaniLSW14stack, Pisani15voidcounts,
  PisaniSW15vpec}.}
With the tracers being (for these
purposes) essentially point objects rather than a continuous
fluid, the Poisson error on an ``average'' galaxy density of,
e.g., 0.2 galaxies in a given number of cubic megaparsecs is
proportionally huge in comparison to studies of overdense
regions. It is thus likely that the debate will require
considerable attention from the community before converging
in regard to optimal strategies for catalogue analysis.
Pisani presented recent work on some of these questions, including
the role of peculiar velocities with respect to an assumed FLRW model,
and predictions for how well void analysis will help in
parametrising cosmological parameters from
the upcoming Euclid and WFIRST missions, provided that the FLRW
model is assumed.

Kaminker presented his work with Ryabinkov \cite{RyabKaminker14} based
on spectroscopic redshifts of 52,683 brightest cluster
galaxies (BCGs) over the interval $0.044 \le z \le 0.78$
from the SDSS DR7 \cite{WenHan13}
and 32,840 {\MgII} absorption line systems (ALSes)
\refpositionsupvsnum{over $0.37 \le z \le 2.28$. \cite{ZhuMenard13}}
{over $0.37 \le z \le 2.28$ \cite{ZhuMenard13}.}
The aim was to analyse structure, separately using
one-dimensional Fourier analysis and two-point auto-correlation functions,
i.e., in effect, projecting the full solid angle of the SDSS into a single pencil beam.
For the $\Lambda$CDM model with $\Ommzero = 0.32$,
the distance ranges of the BCGs range from
130--1900{\hMpc}, while the {\MgII} ALSes lie from
1000--3800{\hMpc} from the observer. Since the diameter of the SDSS
main observing region is roughly a radian, this implies that typically
10--20 BAO peak scale regions cross the observing cone of these two
subsamples. The BAO peak itself is only a small bump above the smooth part
of the two-point auto-correlation function. Unless there is a fair amount
of phase alignment, the tangential projection together of ten to twenty
regions of that size prior to calculation of either a Fourier spectrum
or a correlation function in the radial direction should lead to an
obscuring (convolution) of the bump, leaving a smoothed out version of
the main constituent of the power spectrum of primordial density perturbations.
However, Ryabinkov and Kaminker removed much of the large-scale power by fitting
a smooth function to the redshift distribution for each sub-sample,
so the remaining power should correspond to scales less than a few hundred
megaparsecs.

The results, calculated for a $\Lambda$CDM model with $\Ommzero = 0.25$,
showed an impressively strong (4--5$\sigma$) Fourier signal,
at $98\pm 3${\hMpc} for the BCGs and
at $101\pm 2${\hMpc} for the {\MgII}
\refpositionsupvsnum{ALSes, \cite{RyabKaminker14}}
{ALSes \cite{RyabKaminker14},}
somewhat below the standard BAO peak
scale. Given that typically 10--20 BAO peak scale regions are tangentially mixed together
in each of these sub-samples, this could indicate that the phases of
large-scale structure on the BAO scale in the SDSS are somewhat aligned,
violating the usual assumption of random phases of the primordial density
perturbations. This would be a potentially very interesting test of $\Lambda$CDM,
yielding a much stronger result than the older pencil beam results
that showed a 128{\hMpc}
\refpositionsupvsnum{periodicity, \cite{BroadhurstEKS90Nature}}
{periodicity \cite{BroadhurstEKS90Nature},}
which in
Newtonian structure formation models---that remove small-scale power and introduce a
Lagrangian biasing of the density distribution---was shown to be a natural
\refpositionsupvsnum{outcome, \cite{WeissBuch93}}
{outcome \cite{WeissBuch93},}
while other work using
$N$-body simulations found it to be unlikely at
the 0.1\%
\refpositionsupvsnum{level. \cite{Yoshida01beamsrejected}}
{level \cite{Yoshida01beamsrejected}.}
Full-scale modelling with $N$-body simulations
to see if the Ryabinkov and Kaminker results
are compatible with a standard power spectrum, including BAOs
and random phases,
would be highly justified.

\section{Conclusion}
We have not attempted a Bayesian summary of all available
observational evidence for and against the $\Lambda$CDM
model. It is uncontroversially a good fit to many
observations, despite assuming a prior, Newtonian structure
formation, which has an extremely low (Bayesian) likelihood
compared to general relativity in the only regimes in which
both theories can be directly tested. Pending Bayesian
analyses that include the differences between Newtonian and
relativistic approaches, a frequentist approach to many of
the individual tests that presently reject the $\Lambda$CDM
to high significance, and to those that may potentially
reject it, is justified. The number of disagreements,
including those discussed at MG14---whether labelled
``rejections'', ``inconsistencies'',
``anomalies'' or ``tensions''---cannot be ignored.

Many predictions of the FLRW models are already well
  known, and will continue to be tested in future
astronomical missions such as {\em Euclid}.  In
  contrast, predictions for general-relativistic
  cosmological models that take into account structure
  formation are still in their infancy. However, this field
  of research is developing rapidly, and the initially
  published predictions are likely to be made more robust in
  the coming years.  In particular, rather than seeking
ways to reconcile the $\Lambda$CDM model with
  tests that reject it, we feel that it is important
to think about the observational jigsaw puzzles and come up
with new predictions that are relativistically
  realistic. Relaxing some of the physical restrictions of
the $\Lambda$CDM model, such as a rigid vanishing spatial
curvature on all scales and throughout cosmic history,
may lead to a consensus in the new modelling
efforts that enrich cosmology as a physical science.

\section*{Acknowledgments}

We would like to thank
Mauro Carfora,
George Ellis,
Edward W.~``Rocky'' Kolb,
Malcolm MacCallum,
Jan Ostrowski,
Syksy R\"as\"anen,
Lars Andersson,
Krzysztof Bolejko
and other colleagues at MG14 for constructive
discussions.
The work of TB was conducted within the ``Lyon Institute of Origins'' under grant ANR--10--LABX--66.
TB and BFR acknowledge support for a part of this project from the
HECOLS International Associated Laboratory, supported in part by the
National Science Centre, Poland, grant DEC--2013/08/M/ST9/00664, and
for a part of the project under grant 2014/13/B/ST9/00845 of the
National Science Centre, Poland, and BFR acknowledges visiting
support from CRALyon. AAC acknowledges support from NSERC.

\end{document}